\documentclass[twocolumn,preprintnumbers,amsmath,amssymb,footinbib]{revtex4}

\usepackage{graphicx} 
\usepackage{verbatim} 
\usepackage{dsfont} 
\usepackage[normalem]{ulem}

\begin{document}

\renewcommand{\textfraction}{0.1}

\title{Grand canonical ensemble, multi-particle wave functions, scattering data,\\ and lattice field theories}

\author{Falk Bruckmann}
\affiliation{Universit\"at Regensburg, Institut f\"ur Physik, Universit\"atstra{\ss}e 31, 93053 Regensburg, Germany}

\author{Christof Gattringer}
\affiliation{Universit\"at Graz, Institut f\"ur Physik, Universit\"atsplatz 5, 8010 Graz, Austria}

\author{Thomas Kloiber}
\affiliation{Universit\"at Graz, Institut f\"ur Physik, Universit\"atsplatz 5, 8010 Graz, Austria}

\author{Tin Sulejmanpasic}
\affiliation{North Carolina State University, Department of Physics, Raleigh, NC 27695-8202, USA}

\begin{abstract}
We show that information about scattering data of a quantum field theory can be obtained from studying the system at finite
density and low temperatures. 
In particular we consider models formulated on the lattice which can be exactly dualized to theories of conserved charge fluxes on lattice 
links.  Apart from eliminating the complex action problem at nonzero chemical potential $\mu$, these dualizations allow for a particle world line 
interpretation of the dual fluxes from which one can extract data about the 2-particle wave function.
As an example we perform dual 
Monte Carlo simulations of the 2-dimensional O(3) model at nonzero $\mu$ and finite volume, whose non-perturbative spectrum consists 
of a massive triplet of particles. At nonzero $\mu$ particles are induced in the system, which at sufficiently low temperature 
give rise to sectors of fixed particle number. We show that the scattering phase shifts can be obtained either from the critical chemical potential values separating the sectors or directly from the wave function in the 2-particle 
sector. We find that both methods give excellent  agreement with the exact result. We discuss the applicability and generality of the new 
approaches.
\end{abstract}

\maketitle

In many quantum field theories the low energy excitations are very different from the field content of the elementary Hamiltonian 
or Lagrangian. Indeed excitations, rather than fundamental degrees of freedom, are responsible for 
many important phenomena, ranging from superconductivity in condensed matter to confinement and dynamical mass generation in Quantum 
Chromodynamics (QCD). These excitations are often referred to as particles (or quasi-particles in condensed matter literature) with particular 
masses, charges and interactions, although their internal structure may be rather complicated. For a satisfactory theoretical description 
non-perturbative methods are needed, often complemented with numerical simulations. 

\enlargethispage{\baselineskip}

In the approach discussed here
we consider lattice field theories with a chemical potential $\mu$. In recent years there was quite some progress with finding 
so-called dual representations (see \cite{Chandrasekharan:2008gp,Gattringer:2014nxa} for reviews), 
which provide an exact mapping to new (dual) degrees of freedom, 
which are fluxes for matter and surfaces for gauge fields.
Initially dualization was motivated to overcome the complex 
action problem (at nonzero $\mu$ the action $S$ is complex and the Boltzmann factor $e^{-S}$ cannot be used as a probability in a
Monte Carlo simulation). Here we show that
%
very precise information about 
scattering phases and interactions can be obtained based on two simple observations: 
\begin{enumerate}
\item At finite volume and zero temperature the one, two, three, etc.\ charge sectors are separated by finite energy steps, and hence these charges will appear in the system at certain chemical potential thresholds $\mu_1,\mu_2,\mu_3,...$ the difference of which contains information about the interaction energy of particles carrying charges. 
\item The dual variables allow a particle world line interpretation from which the multi-particle ground state wave function can be obtained. The latter in principle contains all the information about the interactions. 
\end{enumerate}
It is these two observations which are the cornerstone of the two methods we develop in this work: the {\sl charge condensation method} and the {\sl dual wave function method}, respectively.

We develop the ideas using the 2-dimensional O(3) model, and discuss their generality in the end. In addition to being an 
important condensed matter system, the O(3) model is 
a widely used toy theory for QCD because of common features such as asymptotic freedom, a dynamically generated mass gap, and
topological excitations. Moreover, there exist exact results \cite{Zamolodchikov:1977nu} for the scattering phase shifts, i.e., the physical observables
which we aim at in our new approach. Suitable dual representations at nonzero $\mu$ are known 
\cite{Bruckmann:2015sua,Bruckmann:2014sla} and basics
of the finite density behavior were discussed for the related case of the O(2) model in \cite{Banerjee:2010kc}. 

The Lagrangian of a quantum field theory typically shows some continuous symmetries resulting in corresponding conserved Noether charges.
These symmetries and charges play a two-fold role in our approach: When transforming 
the theory to the dual representation the symmetries give rise to constraints of the dual variables making flux conservation explicit. The resulting 
``world lines of flux'' transport the corresponding charge in space and time, and the particle number turns into a topological 
invariant: the winding number of the corresponding flux around the compact Euclidean time. 

The second important aspect of the symmetries and the corresponding charges is that a chemical potential can be coupled to the associated charge, which 
then can be used to populate the corresponding charge sector. In the dual representation the chemical potentials introduce an additional 
weight for the winding number of the corresponding flux lines. 

At sufficiently low temperature $T$ and finite volume one can control the 
particle number with the chemical potential and systematically probe the system in the different charge sectors, 
which are separated by the aforementioned critical values
$\mu_i, i = 1,2, 3 ...$ of the chemical potential. The critical values $\mu_i$ are clearly marked by steps when plotting the expectation value 
of the charge $Q$ at low $T$ as a function of $\mu$ (see \cite{Banerjee:2010kc} and Fig.~\ref{fig:donau} below). 
In the presence of a mass gap $m$, 
the (`Silver blaze') range from $\mu = 0$ to $\mu = \mu_1$ corresponds to the charge-0 sector, no particle is present, and for 
sufficiently large spatial volume one has $\mu_1 = m$.
The interval $\mu\in(\mu_1,\mu_2)$ delimits the charge-1 sector, where $\mu_2$ corresponds to the energy of the lowest charge-2 state, 
i.e., containing two particle masses plus their interaction energy \footnote{The interpretation of the second critical $\mu$ differs if the system contains particles of different charge, as is the case in e.g.\ QCD with gauge group $G_2$ \cite{Wellegehausen:2013cya}.}. At finite volume, 
it is exactly this  2-particle energy that can be related to scattering data using the L\"uscher formula 
\cite{Luscher:1986pf,Luscher:1990ux,Luscher:1990ck}, and analyzing the 2-particle condensation threshold $\mu_2$ as a function of the volume 
constitutes our first approach to extract scattering data from nonzero density and finite volume. We refer to this approach as the 
{\sl ``charge condensation method''}.  

Our second approach -- which we refer to as  the {\sl ``dual wave function method''} -- 
is based on a direct analysis of the dual fluxes in the 2-particle sector, i.e., the interval $\mu\in(\mu_2,\mu_3)$, where 
$\mu_3$ marks the onset of the charge-3 sector. In this interval we analyze the two winding flux loops that characterize the
charge-2 sector and determine the distribution of their spatial distance which we relate to the 2-particle wave function, from which we
again compute scattering data. 

Both methods are presented in detail for the 2-D O(3) model.   
We begin with discussing the dual representation as derived in \cite{Bruckmann:2015sua} where 
the details of the dualization and also the conventional form of the 
model are presented. We consider the O(3) model with a single chemical potential coupled to one of the conserved charges.
The conventional degrees of freedom are O(3) rotors and the chemical potential is coupled to the 
3-component of the corresponding angular momentum. 
In the dual form the partition sum $Z$ is exactly rewritten into a sum over 
configurations $\{m,l,k\}$ of 3 sets of dual variables $m_{x,\nu} \in \mathds{Z}$, $l_{x,\nu} , k_{x,\nu} \in \mathds{N}$ on the 
links $(x,\nu)$ of a $N_t \times N_s$ lattice (periodic boundary conditions, lattice constant $a$):
\begin{eqnarray}
\label{dualz}
Z(\mu) & = & \sum_{\{m,l,k\!\}}  B_J[m,l,k] \; \;
e^{-a\mu\sum_x \! m_{x,0}} 
\\
&& \qquad \times
\prod_x  \, \delta\Big(\sum_\nu[m_{x,\nu}-m_{x-\hat{\nu},\nu}]\Big).
\nonumber
\end{eqnarray}
Each configuration of the variables $m_{x,\nu}, l_{x,\nu}, k_{x,\nu}$ comes with a real non-negative weight $B_J[m,l,k]$ which depends on the
coupling $J$. $B_J[m,l,k]$ is given explicitly 
in \cite{Bruckmann:2015sua}, but irrelevant for the discussion here. A second weight factor 
$e^{-a\mu\sum_x \! m_{x,0}}$ originates from the chemical
potential $\mu$ which couples to the temporal component of the current $m_{x,\nu}$ 
(in our notation $\nu = 0,1$ and $\nu = 0$ denotes Euclidean time).  Obviously all weights are real and positive such that the complex action 
problem is solved in the dual formulation for arbitrary $\mu$.

The dual variables $m_{x,\nu}$, which correspond to one of the O(2) subgroups of O(3), obey constraints at each site $x$. The product of Kronecker deltas $\delta(..)$ enforces
$\nabla \vec{m}_x \equiv \sum_\nu[m_{x,\nu}-m_{x-\hat{\nu},\nu}] = 0 \;  \forall x$, which is a discrete version of a vanishing divergence condition. 
Thus the flux $m_{x,\nu}$ is conserved and the admissible 
configurations of $m_\nu$-flux are closed loops. 

Since the $m_\nu$-flux must form loops, the term in (\ref{dualz}) 
that multiplies $\mu$ can be written as $a\sum_x \! m_{x,0} = a N_t \, w[m_\nu]$, where $w[m_\nu]$ is the total winding number of 
$m_\nu$-flux around the compact time direction with extent $a N_t \equiv 1/T$  (we use natural units
with $k_B = \hbar = c = 1$). Thus we identify the winding number $w[m_\nu]$ as the particle number in the dual formulation. 

\begin{figure}[t]
\includegraphics[width=\linewidth,type=pdf,ext=.pdf,read=.pdf]{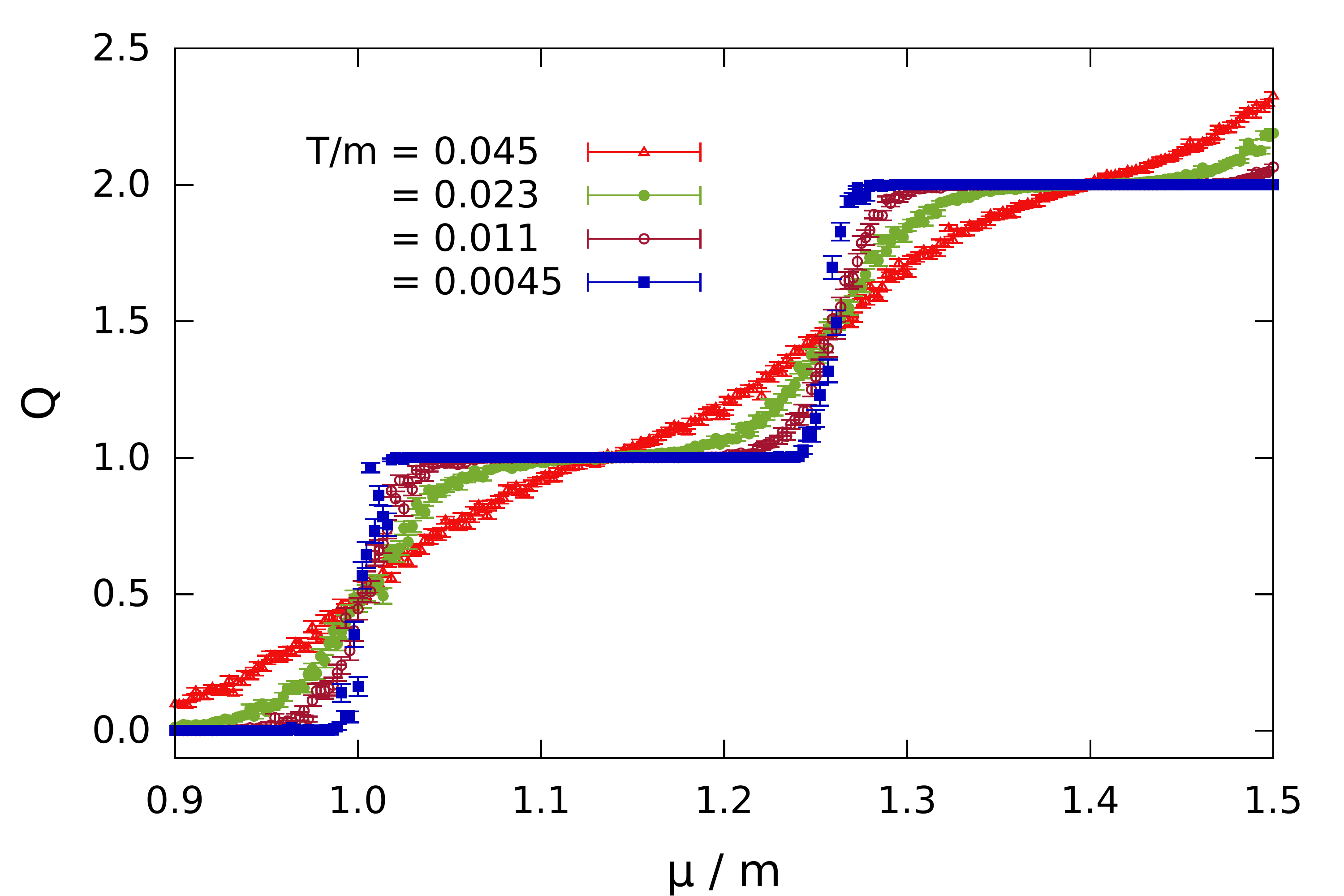}
\caption{The charge $Q$ as a function of $\mu$ (in units of the mass $m$). 
We compare different temperatures $T$ at fixed spatial extent $Lm = 4.4$ ($N_t = 100, 200, 400, 1000$ at $N_s = 20$ using $J = 1.3$).}
 \label{fig:donau}
\end{figure}

\begin{figure*}[!t]
\includegraphics[width=0.48\linewidth,type=pdf,ext=.pdf,read=.pdf]{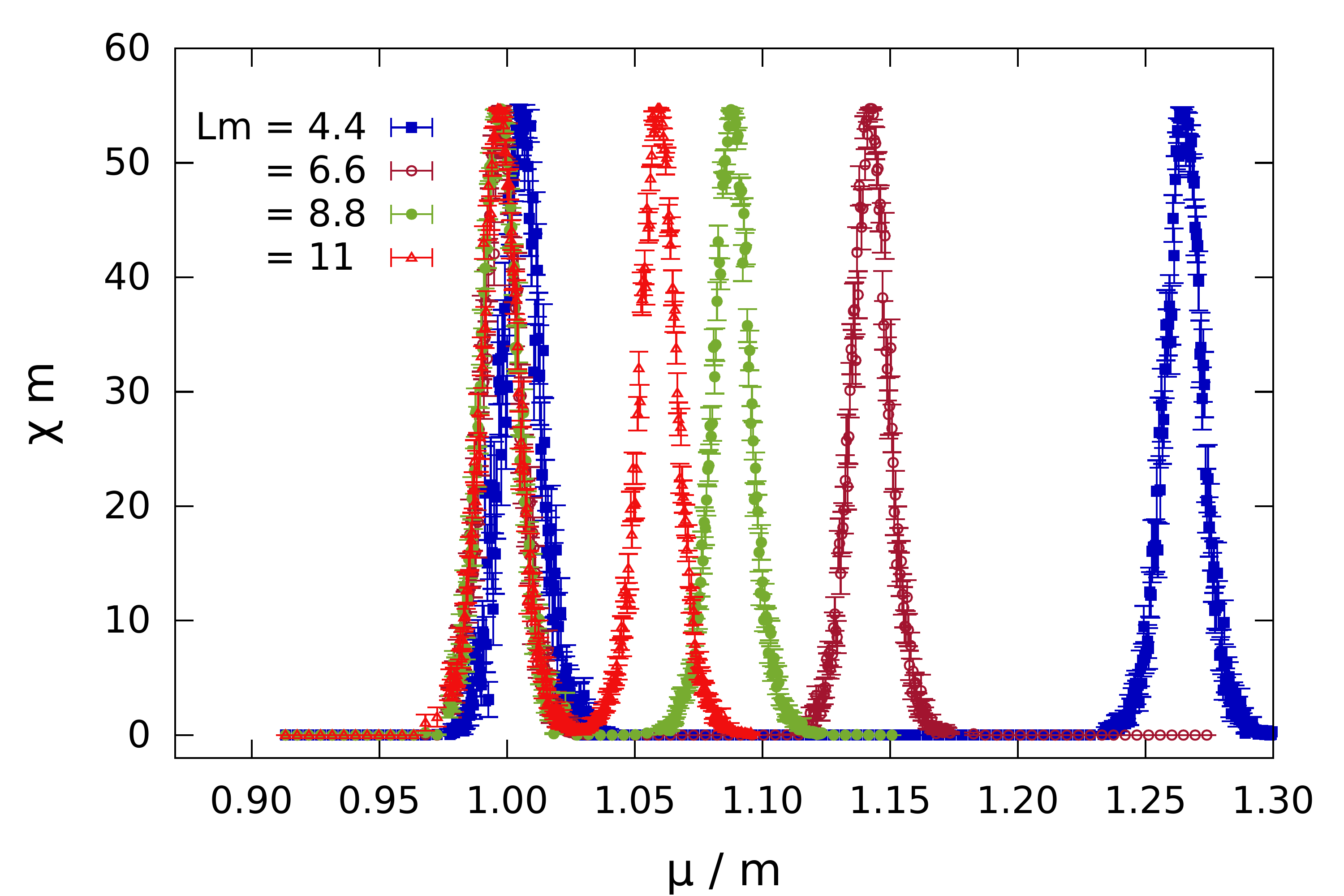}
\hspace{4mm}
\includegraphics[width=0.48\linewidth,type=pdf,ext=.pdf,read=.pdf]{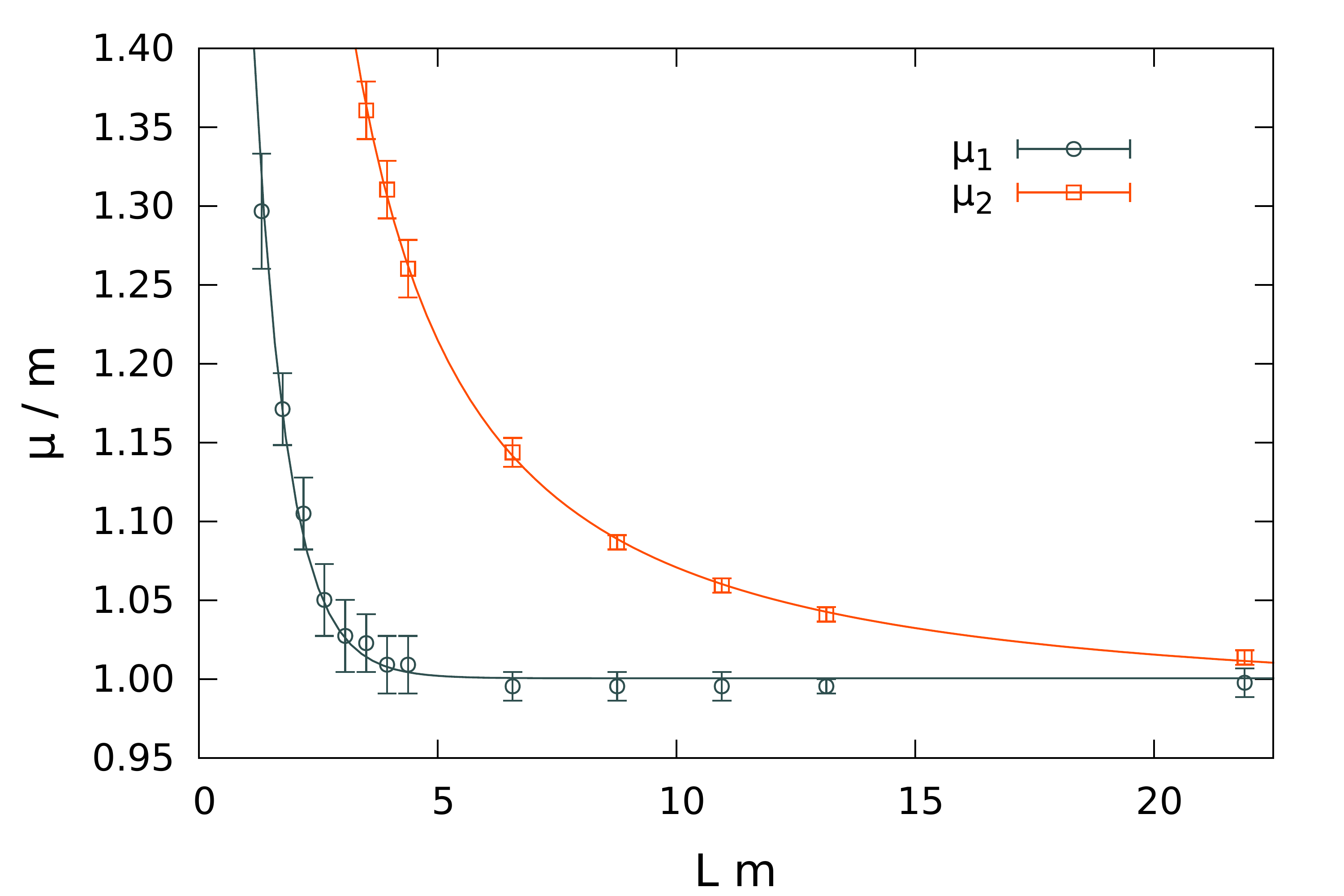}
\caption{Left: The susceptibility  $\chi$ at fixed temperature $T/m=0.0045$ 
as a function of chemical potential for different spatial extensions $L$ ($N_t = 1000$ with $N_s =  20,30,40$ and 50). 
For each $L$ we show the first two peaks corresponding to the critical chemical potentials 
$\mu_{1}$ and $\mu_2$.
Right: The critical chemical potentials at the same temperature now as a function of $L$. The solid curves are results of fits (exponential for $\mu_1$ and power-law for $\mu_2$).}
\label{fig:inn_mur}
\end{figure*}

The constrained $m_\nu$-fluxes can be updated with a generalization of the worm algorithm \cite{Prokofev:2001zz}, while for the other fluxes local
Monte Carlo updates are sufficient. Most of the Monte Carlo results presented in this letter were computed at fixed coupling 
$J = 1.3$ (with some scaling checks performed also at $J = 1.4$ and $J = 1.5$, i.e., closer to the continuum). The temperature 
$T = 1/aN_t$ was varied by changing the temporal extent $N_t$ of the lattice, the spatial size $L = a N_s$ by changing $N_s$.  Dimensionful
quantities are expressed in units of the mass $m$ of the lowest excitation, which was determined from propagators in the conventional 
representation, and at, e.g., $J = 1.3$  is $am = 0.222$.  

To substantiate the above discussion of sectors with fixed $Q$, we show in Fig.~\ref{fig:donau} our $J = 1.3$ 
results for $Q = \langle w[m_\nu] \rangle$ as function of $\mu$ for several low temperatures, 
at fixed $mL = 4.4$. Decreasing the temperature we indeed find the expected formation of plateaus in Fig.~\ref{fig:donau},
which correspond to the sectors of fixed charge, and we can read off the critical values $\mu_i$. In a practical
calculation one actually determines the $\mu_i$ from the peaks of the corresponding particle number susceptibility 
$\chi = \langle (w[m_\nu] - Q)^2 \rangle/L$. These susceptibilities are shown as a function of $\mu$ in the left panel of Fig.~\ref{fig:inn_mur}, 
and we compare results for different spatial extents $L$ in units of $m$. The susceptibilities show pronounced peaks which we can use to determine the values $\mu_i$. We remark that the position $\mu_1$ 
of the first peak is independent of $L$, while the second peak $\mu_2$ shifts to smaller values when increasing $L$ (see the discussion below).

We now make quantitative the above arguments connecting the critical chemical potentials $\mu_1$ and $\mu_2$ with the mass $m$ 
of the lightest particle and the energy of the 2-particle states. We write the grand-canonical partition sum $Z(\mu)$ 
with a grand potential $\Omega(\mu)$ in the form  ($\hat{H}$ and $\hat{Q}$ denote Hamiltonian and charge operator)
\begin{equation}
 Z(\mu) \; =\; \text{tr }e^{-(\hat{H}-\mu\, \hat{Q})/T} \; \equiv \; e^{-\Omega(\mu)/T} \, .
\end{equation}
The small-$T$ limit is governed by the minimal exponents, i.e., in each sector with charge $Q$ the corresponding 
minimal energy $E_{\text{min}}^{Q}$ dominates and the grand potential in the 
different sectors is
\begin{equation}
 \Omega(\mu)\stackrel{T\to 0}{\longrightarrow}
 \left\{
 \begin{array}{ll}
 E_{\text{min}}^{(Q=0)} = 0 & \; \mbox{for} \; \mu\in[0,\mu_1) \; ,\\
 E_{\text{min}}^{(Q=1)}-\mu = m-\mu & \;  \mbox{for} \; \mu\in(\mu_1,\mu_2) \; , \\
 E_{\text{min}}^{(Q=2)}-2\mu = W-2\mu &  \; \mbox{for} \; \mu\in(\mu_2,\mu_3) \; , \\
  \ldots \; \; ,& 
\end{array}
\right.
\label{omega}
\end{equation}
where we introduced the minimal 2-particle energy $W$. Note that in the charge-1 sector we assumed $\mu_1 = m$, i.e., we neglected 
possible finite volume corrections (see below for a cross check).

The 2-particle energy $W$ can be calculated from $\mu_1$ and $\mu_2$:
The first two transitions between neighboring sectors occur when $0=m-\mu_1$ and $m-\mu_2=W-2\mu_2$, respectively, and thus we find
\begin{equation}
 W \; = \; m+\mu_2 \; = \; \mu_1+\mu_2 \; .
 \label{eq:arber}
\end{equation}

So far we have not discussed the role of nonzero spatial momenta of the states, which for our 2-dimensional model are 
given by $2\pi n/L = 2\pi n /a N_s, \, n = 0,1 \, ...\, N_s\!-\!1$. We stress again that at very low $T$
only states with vanishing total momentum contribute to the partition sum. 
Thus in the charge-1 sector we have one particle at rest, in the charge-2 sector two particles with opposite
spatial momentum et cetera. We can combine this with the discreteness of the momenta to understand the $L$-dependence of the $\mu_i$,
which is documented in the right panel of Fig.~\ref{fig:inn_mur}, where we plot the values of $\mu_1$ and $\mu_2$ as a function of $L$. 

For $\mu_1$ we expect a dependence on $L$ only when $L$ becomes smaller than the Compton wave length of the lightest excitation \footnote{The QCD-like theories at nonzero $\mu$ with $L\Lambda_{\text{QCD}}\ll 1$ studied in \cite{Hands:2010zp,Hands:2010vw} are in that regime (and perturbative).} and
self interactions around periodic space alter the mass. This is indeed what we observe in the right panel of Fig.~\ref{fig:inn_mur}.
For $\mu_2$ a-nonvanishing dependence on $L$ is expected throughout, since the box size $L$ controls the allowed relative momentum. Also such a non-trivial dependence is obvious from Fig.~\ref{fig:inn_mur}.
  
\begin{figure}[!t]
 \includegraphics[width=\linewidth,type=pdf,ext=.pdf,read=.pdf]{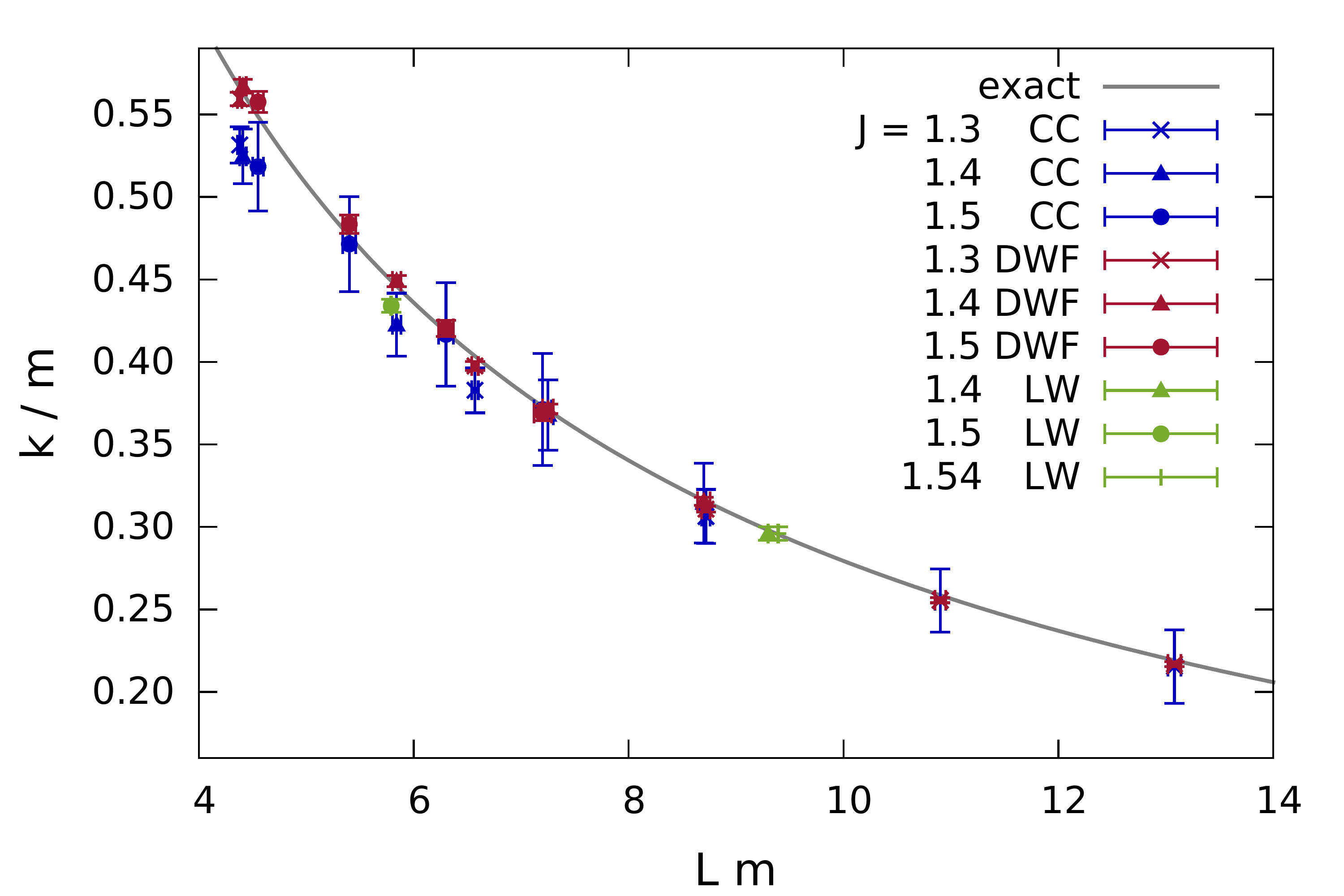}\\
 \includegraphics[width=\linewidth,type=pdf,ext=.pdf,read=.pdf]{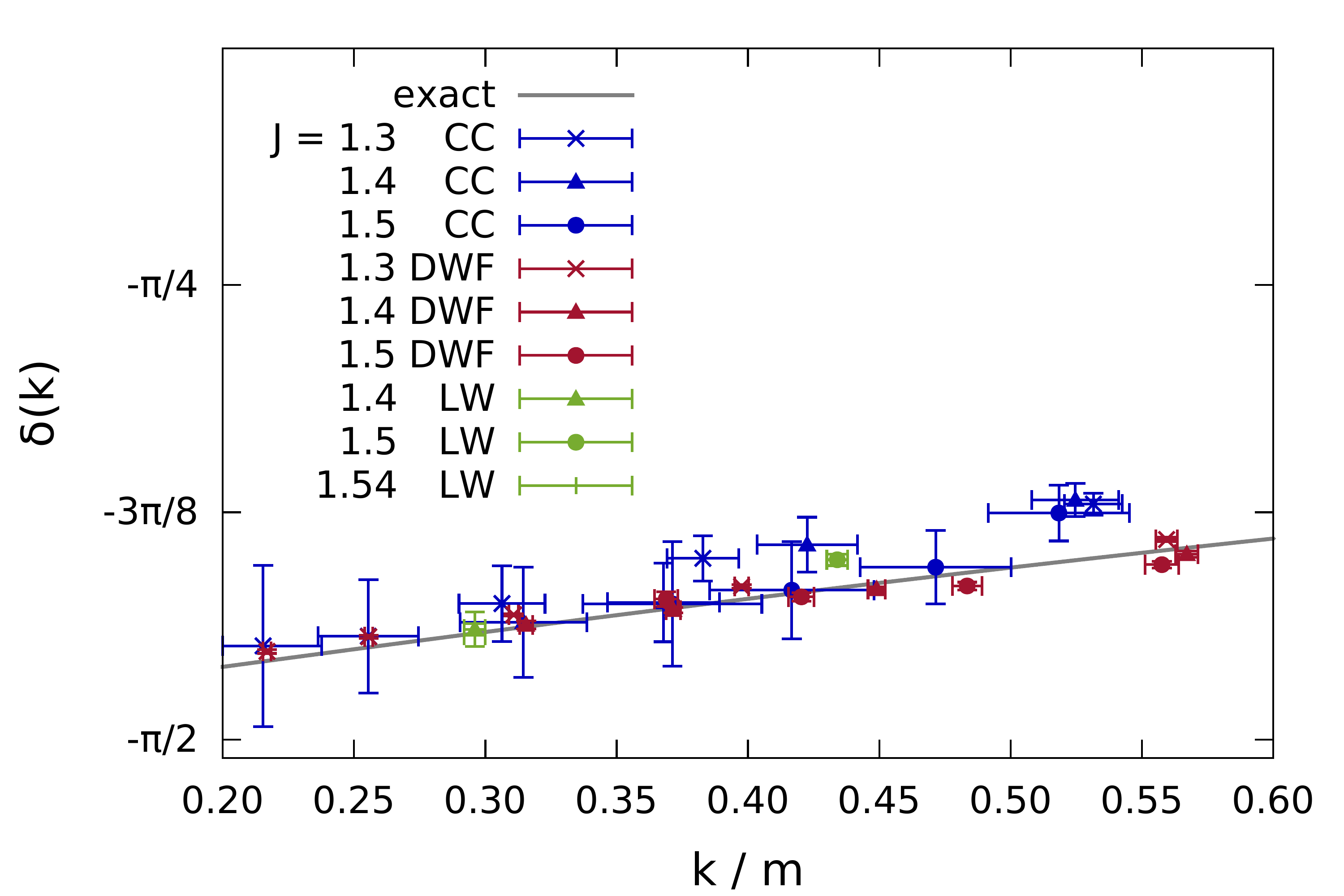}
 \caption{Results from the {\sl charge condensation method} (blue symbols, labelled as CC) and the {\sl dual wave function method} (red, DWF) using 
 different $J$ and $N_s$. We compare our results to the analytic solution \cite{Zamolodchikov:1977nu} (full curves) and the 
 analysis based on 2-particle spectroscopy \cite{Luscher:1990ck} (blue, LW). In the top panel we show the momenta $k$ as a function 
 of $L$ and in the bottom panel the phase shift $\delta(k)$ versus $k$ ($k$ and $L$ in units of $m$).}
\label{fig:isar_elbe}
\end{figure}

For a short ranged potential one can write the two particle energy $W$ as twice the
energy of an asymptotically free particle, 
\begin{equation}
W \; = \; 2\sqrt{m^2+k^2} \; ,
\label{eq:rachel}
\end{equation}
where $k$ denotes the relative momentum which is shifted from the values $2\pi n/L$ of the free case. In a finite volume, only certain quantized values of $k$ can account for the scattering phase shift $\delta(k)$ of the interaction and the periodicity, as expressed by the L\"uscher formula \cite{Luscher:1986pf} 
\begin{equation}
e^{2 i\delta(k)} \; = \; e^{-ikL} \; .
\label{eq:lusen}
\end{equation}
Varying $L$ allows one to scan a whole range of momenta $k$. The temperature must be low enough for pronounced plateaus to form (see Fig.~\ref{fig:donau}), which gives rise to the following two conditions: $T\ll m$ and $T/m\ll 1/(Lm)^2$. Our numerical results for this extraction are given in Fig.~\ref{fig:isar_elbe}.
In the top panel we show our data for $k$ as a function of $L$ and compare to the exact result \cite{Zamolodchikov:1977nu} for ``isospin 2'', which is the  
relevant case for our choice of the chemical potential which excites the 3-component of the O(3) angular momentum. 
We also include  
results from the numerical spectroscopy 
calculation in the 2-particle channel \cite{Luscher:1990ck} and find excellent agreement of our results with the analytical and numerical reference
data. In the bottom panel we give the results for the phase shift $\delta(k)$ as a function of $k$, and again find excellent agreement  with the reference 
data.
 
Our second approach for the determination of scattering data, 
the {\sl ``dual wave function method''}, 
is based on a direct analysis of the flux variables $m_{x,\nu}$ which carry the charge: 
In the charge-2 sector we identify the two flux lines which wind around the compact time
and interpret their temporal flux segments $m_{x=(x_0,x_1),\nu=0}=1$ as the spatial position $x_1$ 
of the charge at time $x_0$. 
Thus for a given time $x_0$ we obtain two positions $x_1^{(1)}$, $x_1^{(2)}$ and identify 
$\Delta x = | (x_1^{(1)}-x_1^{(2)}) |$ as the distance of the two charges at that time. Sampling over time and
many configurations we obtain the probability distribution of the distance $\Delta x$, and the square root of this
distribution can be identified as the relative wave function $\psi(\Delta x)$ of the two charges.   

\begin{figure}[!t]
 \includegraphics[width=\linewidth,type=pdf,ext=.pdf,read=.pdf]{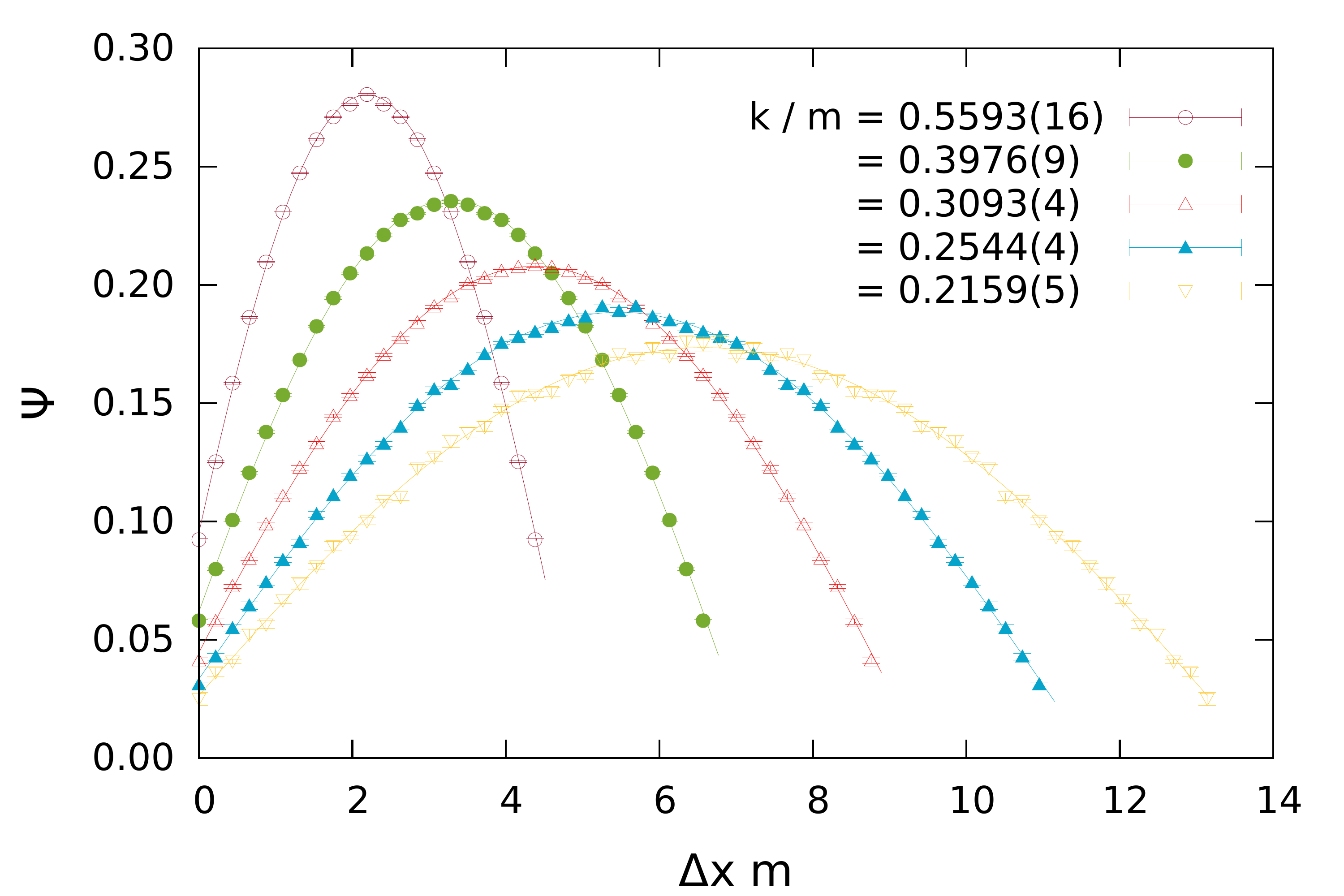}
 \caption{The relative wave function $\psi(\Delta x)$ as determined from analyzing the flux lines in the charge-2 sector. We show results for 
 $J = 1.3$ at $T/m = 0.0045$ $(N_t = 1000)$ and different spatial extents, $Lm=4.4, 6.6, 8.8, 11, 13.2$ ($N_s=20,30,40,50,60$). The wave functions are fit according to Eq.~(\ref{eq:dachstuhl}) 
 and in the legend we give the corresponding fit results for $k$.}
 \label{fig:regen}
\end{figure}

In Fig.~\ref{fig:regen} we show  $\psi(\Delta x)$ for different spatial extents $L$.
Except for very small $\Delta x$ the wave functions are very well described by shifted cosines, 
\begin{equation}
 \psi(\Delta x) \; \propto \; \cos(k(\Delta x-L/2)) \; .
 \label{eq:dachstuhl}
\end{equation}
This finding confirms the applicability of Eq.~(\ref{eq:rachel}): Outside the interaction range the wave functions for the relative motion 
of the two particles are standing waves with momenta $k$ which are related to the scattering phase shifts via (\ref{eq:lusen}).
Thus we can fit the data for $\psi(\Delta x)$ with the cosines (\ref{eq:dachstuhl}) and obtain the momenta $k$ from that fit.  
The fit results are also shown in Fig.~\ref{fig:regen} and in the legend we give the corresponding momenta $k$. From these one again
obtains the phase shift $\delta(k)$ via (\ref{eq:lusen}). The results from the {\sl dual wave function method} are 
included in Fig.~\ref{fig:isar_elbe} and agree very well with the {\sl charge condensation method} 
and the reference data.

Let us summarize the two approaches for obtaining scattering data presented here and comment on their generality: Both methods are 
based on simulations at nonzero chemical potential $\mu$ at very low temperatures.  The {\sl charge condensation method} relates the critical chemical potentials $\mu_1$ and $\mu_2$ to the 2-particle energy, which in turn can be related to the scattering phase shift via the L\"uscher formula. 
Technically this method assumes that nonzero chemical potential simulations are feasible, which for example is the case for the isospin potential in QCD.
The basic concept of the {\sl charge condensation method} is generalizable also to higher dimensions, in particular the relation of the condensation thresholds 
$\mu_1$, $\mu_2$ to the 2-particle energy. In higher dimensions a full partial wave analysis would be necessary for complete 
information about scattering, but at least the scattering length can be extracted from the 2-particle energy 
\cite{Luscher:1986pf,Luscher:1990ux}. 
For the second approach, the {\sl dual wave function method}, the chemical potential is adjusted such that the system is
in the charge-2 sector. This method is rather general in arbitrary dimensions. 
The relative 2-particle wave function can be determined from the fluxes in the charge-2 sector and in principle this gives access to the 
complete scattering information. 

We remark that both methods can be straightforwardly generalized to three and more particles. A second remark concerns the possibility to couple different chemical potentials to different
conserved charges. Using such a setting one can also populate 2-particle sectors with two different particle species and study 
their scattering properties. 
While the {\sl charge condensation method} is independent of the representation of the system, we emphasize that the {\sl dual wave function method} requires a suitable dualization and makes explicit use of the wordline interpretation. The latter method opens the possibility to study particles and antiparticles, i.e., charges of opposite sign.

This exploratory paper is a first step towards a full understanding of the relation 
between low temperature multi-particle sectors and scattering data. We believe that this is an interesting connection to explore, and 
understanding the physics involved clearly goes beyond the development of a new method for determining scattering data in lattice simulations.


\begin{acknowledgments}
FB is supported by the DFG (BR 2872/6-1) and TK by the Austrian Science Fund, 
FWF, DK {\sl Hadrons in Vacuum, Nuclei, and Stars} (FWF DK W1203-N16).  
Furthermore this work is partly supported by the Austrian Science Fund FWF Grant I 1452-N27 and by  
DFG TR55, {\sl ``Hadron Properties from Lattice QCD''}. We thank G.\ Bali, J.\ Bloch,  
G.\ Endr\H{o}di, C.B.\ Lang, J.\ Myers, A.\ Sch\"afer and T.\ Sch\"afer for discussions. 
\end{acknowledgments}

\bibliography{bibliography}

\end{document}